\documentclass[fleqn,twoside]{article}
\usepackage{espcrc2}

\usepackage{psfig}
\usepackage[figuresright]{rotating}

\newcommand{\AmS}{{\protect\the\textfont2
  A\kern-.1667em\lower.5ex\hbox{M}\kern-.125emS}}

\title{SYM Correlators and the Maldacena Conjecture}

\author{Uwe Trittmann, Department of Physics, Ohio State University, 
	Columbus, OH        
        \thanks{Work in collaboration with S.S.~Pinsky and J.R.~Hiller.}}
       
\begin{document}
\begin{abstract}
We report on progress 
in evaluating quantum filed theories with 
supersymmetric discrete light-cone quantization (SDLCQ).
We compare the method to lattice gauge theory and point out its relevance 
for lattice calculations.
As an exciting application we present a test of the Maldacena conjecture.
We test the conjecture by
evaluating the correlator of the stress-energy tensor in the
strong coupling field theory and comparing to the string 
theory prediction of its behavior as a function of the distance.
Our numerical results support
the Maldacena conjecture and are within 10-15\% of the predicted results.
\vspace{1pc}
\end{abstract}

\maketitle


\section{Introducing the Method: SDLCQ}

Supersymmetric Discretized 
L{ight-}C{one} Quantization (SDLCQ)
is a discrete, Hamiltonian, manifestly supersymmetric 
approach to solving quantum field theories. 
Light-cone coordinates $(x^+,x^-,\vec{x}^\perp)$ are defined as 
\begin{equation}
x^\pm=(x^0\pm x^1)/\sqrt{2},
\end{equation} 
where 
${ x^+}({ x^-})$ {plays the role of a time(space) coordinate.
Transverse coordinates are treated in the usual way.
The conjugate variables are 
$(P^+)p^+$, the (total) longitudinal momentum.
The light-cone energy
$P^-$ is the Hamiltonian operator which propagates the system
in the light-cone time, and is of utmost importance.
In light-cone quantization all individual 
longitudinal momenta are positive, $p_i^+\ge 0$.
This allows for a convenient discretization of the 
theory by putting (anti-)periodic boundary conditions
on the fields. The momenta are then characterized by an integer $n_i$
which symbolizes a momentum fraction
\begin{equation}
p_i^+=\frac{n_iP^+}{K};\qquad n=1,2,3,\ldots,K. 
\end{equation}
Here, 
$K{\equiv}P^+{L}/{\pi}$ 
is the harmonic resolution and also by construction 
the maximal number of partons.
The continuum limit is reached as $K\rightarrow \infty$.

The framework of DLCQ can be utilized to create a manifestly
supersymmetric approach, namely SDLCQ. 
The key ingredient is the
preservation of supersymmetry even at finite cutoff by
discretizing {\em first} {the supercharge}
$Q^-$ {and} {\em then} constructing 
{the Hamiltonian} via
\vspace*{-0.05cm}
\begin{equation}
P^-=\frac{1}{2\sqrt{2}}\{Q^-,Q^-\}.
\end{equation}
\vspace*{-0.05cm}

\begin{table*}[htb]
\centerline{
\begin{tabular}{|c|c|}\hline
{\bf SDLCQ} & {\bf Lattice}\\\hline
Hamiltonian formalism & action based \\
solve EVP {$2P^+P^-|\psi_n\rangle=M^2_n|\psi_n\rangle$} & evaluate partition function
{$Z=\int [d\phi] e^{-S}$}\\
discretization parameter {$K$} & lattice spacing {$a$}\\
exact {SUSY} at each step & exact {gauge invariance} at each step \\
get complete spectrum & get (preferably) lowest states\\
computational effort small & computational effort substantial\\
({but:} exponential rise with {$K$})  & \\
{no} 
stochastic methods for evaluation & stochastic methods for evaluation\\
large {$N_c$} 
simplifies calculations & need to extrapolate to large {$N_c$}\\
massless particles no problem & need to extrapolate to 
{$m\rightarrow 0$}\\
SUSY essential ingredient & SUSY problematic (massless fermions)\\ 
\hline
\end{tabular}}
\caption{A schematic comparison of SDLCQ and lattice gauge theory.\label{table1}}
\end{table*}

A schematic comparison between the essential properties of SDLCQ and lattice
gauge theory is compiled in Table \ref{table1}.
Since the approaches are complementary, results can be tested against 
each other!
Interesting results in this direction have been obtained by
Hamiltonian lattice methods \cite{Campostrini}
and in work on supersymmetry on the lattice \cite{Montvay}.
There is a host of results in SDLCQ in two and three dimensions
on correlators \cite{correlators}, bound states \cite{masses}, 
and other topics, including an overview article, Ref.~\cite{Overview}.


\section{Application: Maldacena conjecture}

The Maldacena conjecture \cite{Maldacena} states, {\em cum grano salis},
that a field theory can be equivalent to a 
string theory on a special background.
The drawback of the exciting perspectives of this conjecture 
are the problems to verify it.
The crucial issue is that we need
a matching point where the theories are equivalent. It should have	
small curvature, so that the supergravity  approximation is valid,
together with a small coupling allow for the use of perturbation theory.
There is no such scenario known.
Here the non-perturbative features of SDLCQ come to the rescue.

A variant of the Maldacena conjecture states that two-dimensional 
${\cal N}=(8,8)$ supersymmetric Yang-Mills (SYM) theory 
should be equivalent to a system 
of D1 branes in Type IIB string theory decoupling from gravity 
\cite{Itzhaki}.
We will use the correlation function of a gauge invariant operator, namely 
{$T^{\mu\nu}$}, to test this conjecture.
The agenda is then clear: 
we have to compute the form of correlator in supergravity (SUGRA) 
approximation, and then
perform a non-perturbative calculation of the correlator in {SDLCQ}.


\subsection{The Correlator from SUGRA}
\label{Sec21}

One can compute the two-point correlation function of the 
stress-energy tensor from
string theory using the SUGRA (i.e.~small curvature) approximation 
\cite{ItzhakiHashimoto}
The leading non-analytic term in the flux factor yields the correlator
\begin{equation}
{\langle {\cal O}(r){\cal O}(0)\rangle=}{{N_c^{3/2}/}{g r^5}}.
\end{equation}
As a check we remark that {${\cal N}{=}(8,8)$ SYM$_2$} 
has conformal fixed points at the ultra-violet and the infra-red
with central charges {$N_c^2$} and {$N_c$}, respectively. 
One expects to deviate from the conformal behavior at 
$r_{UV}{=}(g\sqrt{N_c})^{-1}$,
and  $r_{IR}{=}{\sqrt{N_c}}g^{-1}$.
This yields the following phase diagram: 
\bigskip

\centerline{
\unitlength0.5cm
\begin{picture}(15,2)\thicklines
\put(0.2,1){\vector(1,0){14.8}}
\put(0.2,0.8){\line(0,1){0.4}}
\put(5,0.8){\line(0,1){0.4}}
\put(10,0.8){\line(0,1){0.4}}
\put(0,0){$0$}
\put(4.5,0){$\frac{1}{g\sqrt{N_c}}$}
\put(9.5,0){$\frac{\sqrt{N_c}}{g}$}
\put(14.8,0.2){$r$}
\put(2,0.3){UV}
\put(6.5,0.3){SUGRA}
\put(12,0.3){IR}
\put(1.5,1.5){$N_c^2/r^4$}
\put(6.0,1.5){$N_c^{3/2}/(g r^5)$}
\put(11.5,1.5){$N_c/r^4$}
\end{picture}
}


\subsection{The correlator from SDLCQ}

To reproduce SUGRA scaling relation, we will calculate the 
cross-over behavior at  $1/g\sqrt{N}$}{$<r<$}{$\sqrt{N}/g$} 
using {SDLCQ}.
We want to compute correlator
{
\begin{equation}
F(x^-,x^+)=\langle{\cal O}(x^-,x^+){\cal O}(0,0)\rangle.
\end{equation}}
As a gauge invariant (two-body) operator we take {$T^{++}(-K)$}.
In {DLCQ} one fixes {$P^+={K\pi}/{L}$}. Therefore we 
Fourier transform the last equation and decompose it into modes.
We then continue to Euclidean space by taking distance 
{$r^2 = 2 x^+ x^-$} to be real. This yields
\begin{eqnarray}
{\cal F}(r)&=&\left(\frac{x^{-}}{x^+}\right)^2 F(x^-,x^+)\label{corr}\\
&=&\sum_n
\underbrace{\left|{L \over \pi} \langle n | T^{++}(-K) |0 \rangle \right|^2}_
{\mbox{matrix element independent of L}}\nonumber\\
&&\times
\underbrace{{M_n^4 \over 8 \pi^2 K^3} {\cal K}_4(M_n r).}_{\mbox{function of eigenvalues}}\nonumber
\end{eqnarray}
We emphasize that this result is 
dependent on the harmonic resolution $K$, 
but involves no other unphysical quantities.
We recover the continuum limit by sending {$K\rightarrow\infty$}.
The correct small {$r$} behavior is retained. 


From the numerical perspective the evaluation of 
the expression for  {${\cal F}(r)$} is straightforward. 
We need to calculate the mass spectrum
by solving eigenvalue problem, {\em i.e.}~diagonalizing the Hamiltonian, 
\begin{equation}
 2P^+P^-(K)|\psi_n(K)\rangle=M^2_n(K)|\psi_n(K)\rangle.
\end{equation}
The problem is the large number of particles in the theory which has the 
Fock space growing exponentially with 
the harmonic resolution {$K$}.
The necessary numerical improvements include
writing a {C++} code with an efficient data structure, incorporation of
the discrete flavor symmetry of the problem, and an increase of 
numerical efficiency by an improved version of the 
{Lanzcos algorithm}.
The hardware requirements are quite modest. We work with a  
Linux workstation with a  Pentium III processor at 733 MHz and 2 GB RAM.
Typical running times for large-scale computations are in the order of a few days.

\subsection{Results}

The correlator {${\cal F}(r)$}, Eq.~(\ref{corr}), is 
determined by a numerical calculation of
the mass spectrum of the ${\cal N}=(8,8)$ SYM theory. 
One problem with the discrete approach is the existence of unphysical states. 
Additionally, the number of partons in the massless unphysical 
states is {even}/{odd} for {$K$} {even}/{odd}.
Since the correlator, Eq.~(\ref{corr}), is only sensitive to two-particle 
contributions, the resulting curves {${\cal F}(r)$} are different for  
{even} and {odd} {$K$}.
Furthermore, the unphysical states yield also a $1/r^4$ behavior, 
but have a {wrong} and dominant {$N_c$} dependence.
Therefore we cannot see  regular contribution at large {$r$}.
We can, however, take the different behavior of the 
curves to establish where the approximation breaks down.
As a consistency check we note that
the approximation breaks down at larger and larger {$r$} as the harmonic resolution
{$K$} grows.

\begin{figure}
\centerline{
\psfig{file=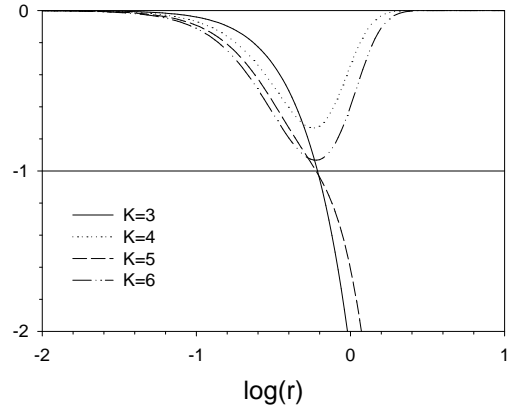,width=7true cm,angle=0}}
\vspace*{-1cm}
\caption{
Log-log
derivative with respect to {$r$} of the correlator
${\cal F}(r)\times
\left({x^- \over x^+} \right)^2 {4 \pi^2 r^4 \over N_c^2 (2 n_b +n_f)}$
vs.~$r$ in units {$g^2 N_c /\pi$} for {$K=3,4,5$} and $6$. \label{fig1}
}
\end{figure}

From the discussion of Sec.~\ref{Sec21}
we expect the 
correlator {${\cal F}(r)$} to change its behavior from  
{$1/r^4$ to $1/r^5$} as {$r$} increases. 
We should thus  approach {$d{\cal F}/dr=-1$} 
in the continuum limit
and would claim success if the derivative 
flattens at this value before the approximation breaks down.
As we see in Fig.~\ref{fig1}, the derivative approaches this
line, but the approximation breaks down when the 
curve reaches a value of $d{\cal F}/dr\approx -0.85$ at $K=6$.
We therefore have evidence that the Maldacena conjecture is correct,
although not yet a decisive result.

\section{Conclusions and Perspectives}
\vspace*{-0.1cm}

To put things in perspective we state that
{SDLCQ} is a viable way to solve quantum field theories.
Results within this framework include spectra, correlators  and 
other properties of two- and three-dimensional YM- and SYM-theories.
A test of these results by an independent method, namely lattice gauge theory,
is necessary and would be very much appreciated!  
As an example we showed a test of the Maldacena conjecture.
Though the results are not totally conclusive, the values
are within {10-15\%} of results expected from the conjecture. 
Improvements of the code and the numerics are possible
and on the way.

\end{document}